\documentclass[preprint, double-spaced,showpacs,preprintnumbers,amsmath,amssymb]{revtex4}

\usepackage{graphicx}
\usepackage{dcolumn}
\usepackage{bm}

\begin{document}


\title{Measurement of the Spin-forbidden Decay rate (3s3d)$^{1}$D$_{2}$ $\rightarrow$ (3s3p)$^{3}$P$_{2,1}$ in $^{24}$Mg} 

\author{K. T. Therkildsen}\email{kaspertt@fys.ku.dk}\author{B. B. Jensen, C. P. Ryder, N. Malossi}
\author{J. W. Thomsen}%

\affiliation{%
The Niels Bohr Institute,
Universitetsparken 5, 2100 Copenhagen, Denmark}%

\date{\today}

\begin{abstract}
We have measured the spin-forbidden decay rate from
(3s3d)$^{1}$D$_{2}$ $\rightarrow$ (3s3p)$^{3}$P$_{2,1}$ in $^{24}$Mg
atoms trapped in a magneto-optical trap. The total decay rate,
summing up both exit channels (3s3p)$^{3}$P$_{1}$ and
(3s3p)$^{3}$P$_{2}$, yields (196 $\pm$ 10) s$^{-1}$ in excellent
agreement with resent relativistic many-body calculations of [S.G.
Porsev et al., Phys. Rev. A. \textbf{64}, 012508 (2001)]. The
characterization of this decay channel is important as it may limit
the performance of quantum optics experiments carried out with this
ladder system as well as two-photon cooling experiments currently
explored in several groups.

\end{abstract}

\pacs{Valid PACS appear here}
\maketitle

\section{\label{Intro}Introduction}
The interest in alkaline earth systems has increased significantly
over the past years. One of the main interests in these systems is
their applications to high resolution spectroscopy and atomic
frequency standards, as they offer very narrow electronic
transitions \cite{Intro}. Another attractive feature of the
two-electron systems is their very simple energy level structure,
for bosonic isotopes, free of both fine and hyperfine structure.
This simple structure opens for detailed comparison between theory
and experiment \cite{Machholm}. Recent advances in \emph{ab initio}
relativistic many-body calculations of two-electron systems have
provided very accurate values for atomic structure and properties
such as the electric-dipole transitions of Mg, Ca, and
Sr\cite{porsev}. However, in the case of Mg relatively few
experiments have been
reported in literature motivating the studies in this Brief Report.\\

In a series of papers two photon cooling in ladder schemes have been
proposed to lower the temperature below the doppler limit of
alkaline earth elements \cite{Cruz,Dunn,Arimondo}. The magnesium and
calcium systems are of particular interest since cooling is limited
to a few mK. Experimental evidence for two-photon cooling has been
established, but the effect is rather modest contrary to theoretical
predictions \cite{Malossi,Mehlstaubler}. One important mechanism to
take into account in these studies are leaks from the ladder system,
limiting the number of atoms that eventually can be cooled.

In this this Brief Report we present the first measurements of the
spin forbidden decay from the magnesium (3s3d)$^{1}$D$_{2}$ state to
(3s3p)$^{3}$P$_{1}$ and (3s3p)$^{3}$P$_{2}$. Using atoms trapped in
a magneto-optical trap we optically prepare the atoms in the
(3s3d)$^{1}$D$_{2}$ state and measure the decay to
(3s3p)$^{3}$P$_{2,1}$ manifold by trap loss measurements. This
enables us to deduce the absolute decay rate for these transitions
and compare our results to state of the art theoretical
calculations.

\section{\label{Theory}Trap loss measurements}
 \begin{figure}[h]
\includegraphics*[width=0.95\columnwidth]{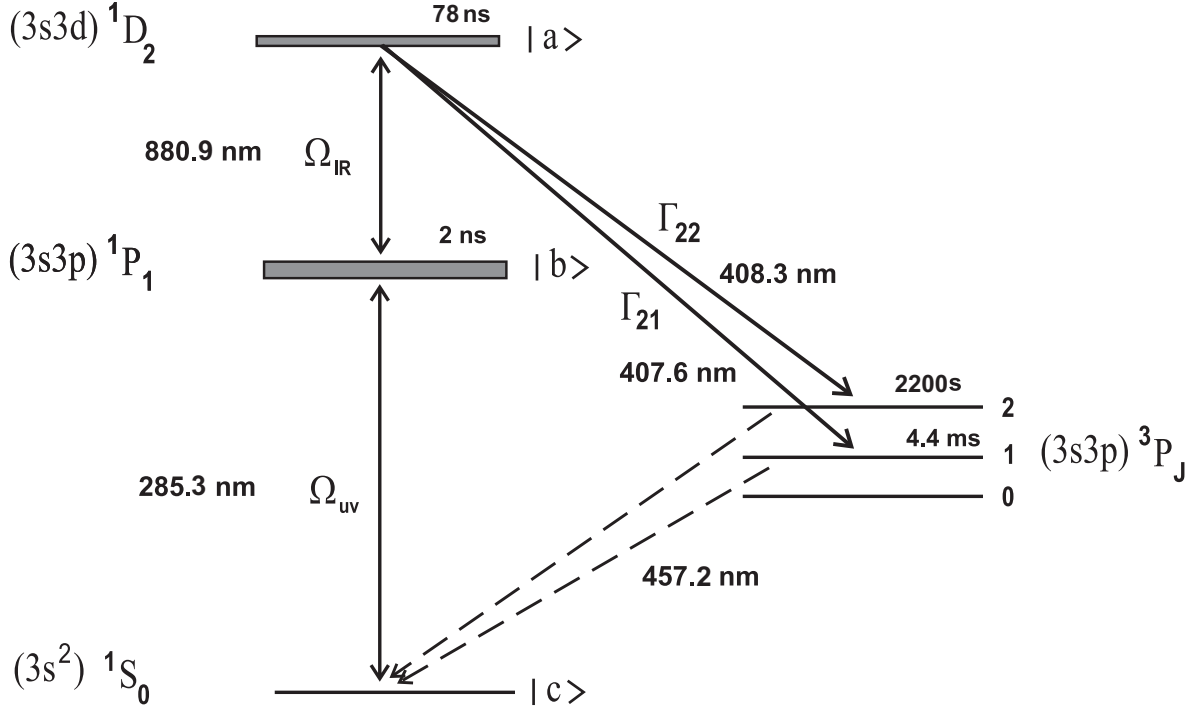}
\caption{Energy levels relevant for the (3s3d)$^{1}$D$_{2}$
$\rightarrow$ (3s3p)$^{3}$P$_{2,1}$ decay measurement in
$^{24}$Mg.}\label{energylevels}
\end{figure}
In figure \ref{energylevels}, we show the relevant energy levels and
transitions in our experiment. The 285 nm transition is used for
coolingand trapping the atoms while the 881 nm transition is used to
populate the $(3s3d)\;^1{\rm D}_2$ state. From this state the atoms
may decay to the (3s3p)$^{3}$P$_{1}$ state  (407.6 nm) with rate
$\Gamma_{21}$ or to the (3s3p)$^{3}$P$_{2}$ state (408.3 nm) with
rate $\Gamma_{22}$, or simply back to (3s3p)$^{1}$P$_{1}$. By
neglecting decay to the (3s3p)$^{3}$P$_{0}$ state, which is weaker
by orders of magnitude, no other decay channels are open. When the
881 nm laser is switched off we express the steady state number of
atoms $N_0$ as the ratio of the load rate to linear losses
$L/\alpha$. Here we neglect intra-MOT collisions since we are
working at relative low atom densities. The equation for the number
of atoms trapped in the MOT with 881 nm light present can be written
as:
\begin{eqnarray}
  \dot{N} &=& L-\rho_{aa}\,(\Gamma_{22}+\Gamma_{21})N-\alpha\,N.
\end{eqnarray}
Experimentally we determine the total loss rate
$\Gamma_{22}+\Gamma_{21}$. In our case the linear loss rate $\alpha$
is dominated by resonant photo-ionization of (3s3p)$^{1}$P$_{1}$ at
285 nm \cite{madsen3}. Photo-ionization from the $(3s3d)\;^1{\rm
D}_2$ level is below 1 Mb and will not be considered
here\cite{fang}. The steady state becomes
\begin{equation}\label{steadystate1}
N^{on}=\frac{L}{\rho_{aa}\,(\Gamma_{22}+\Gamma_{21})+\alpha^{on}},
\end{equation}
and
\begin{equation}\label{steadystate2}
N^{off}=\frac{L}{\alpha^{off}}.
\end{equation}
The ratio of signal with 881 nm turned on and off is proportional
to $\rho_{bb}^{on}$, $\rho_{bb}^{off}$ respectively. We find
finally
\begin{equation}
\frac{S^{on}}{S^{off}}=
\frac{\alpha^{off}\,\rho_{bb}^{on}/\rho_{bb}^{off}}{\rho_{aa}\,\bigg(\Gamma_{21}+\Gamma_{22}\bigg)+\alpha^{off}\,
\frac{\rho_{bb}^{on}}{\rho_{bb}^{off}}},\label{SondivSoff}
 \end{equation}
as the $\alpha$-coefficient is dominated by resonant
photo-ionization and thus proportional to the fraction of atoms in
the (3s3p)$^{1}$P$_{1}$ state. So far we assumed that all atoms
decaying to the (3s3p)$^{3}$P$_{J}$ manifold are lost. This is
indeed the case for the (3s3p)$^{3}$P$_{2}$ final states due to
their long lifetime exceeding 2200 seconds \cite{porsev2}. On the
time scale of the experiment, being only a few seconds, we consider
the atoms from this state lost. However, for the (3s3p)$^{3}$P$_{1}$
states atoms in the $|J=1, m_j=+1 \rangle$ magnetic sub state will
be trapped in the magnetic quadrupole field of the MOT and most
likely be recaptured when decaying back to the (3s$^2$)$^{1}$S$_{0}$
ground state. Atoms in $|J=1, m_j=0,-1 \rangle$ magnetic sub states
are expected to be lost since the rms velocity of the trapped atoms
exceed 1 m/s leading to a rms travelled distance of more than 5 mm,
which is large compared to our MOT beam diameter of only a few mm.
The total decay rate will thus be modified to
$\gamma\Gamma_{21}+\Gamma_{22}$. The actual value of $\gamma$ depend
strongly on the polarization of the 285 nm beam and the 881 nm beam.

\begin{figure}[h]
\includegraphics*[width=0.95\columnwidth]{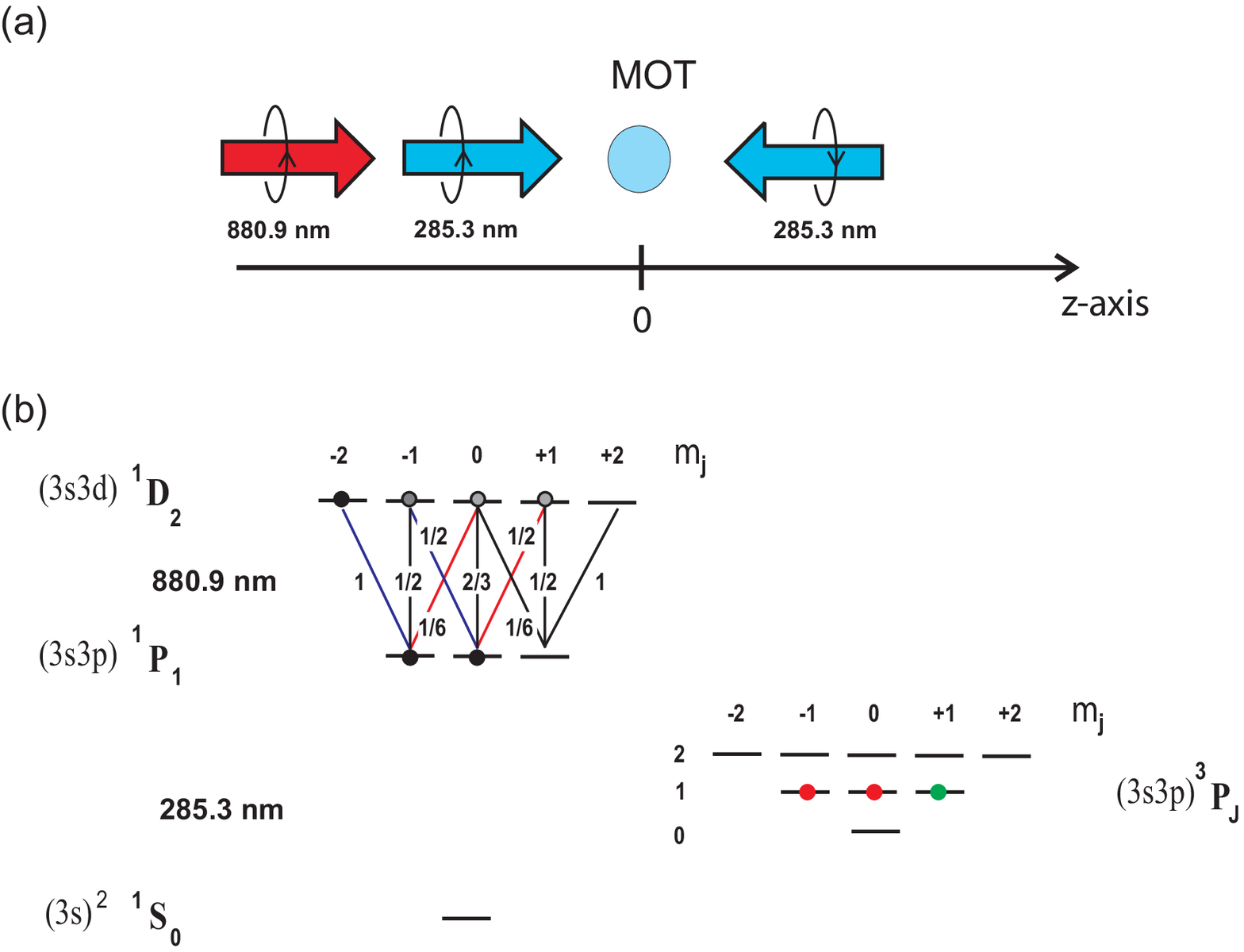}
\caption{ (color online)(a) Experimentally the 881 nm beam is
co-linear with one of the MOT beams. The polarization state can be
selected similar to the MOT helicity, as shown in the figure, or
opposite. (b) Polarization dependent pumping of atoms in the MOT.
Assuming a one dimensional representation of the MOT the 881 nm
polarization, represented by blue line (= MOT helicity) or red line,
optical pumping will drive the atoms such that anisotropy in the
decay rate will emerge. The green dot represents atoms trapped in
the MOT quadrupole magnetic field, while red dots represents atoms
in non-trapped states. The Clebsch Cordan coefficients shown will
also apply to the $(3s3d)\;^1{\rm D}_2$ $\rightarrow$
(3s3p)$^{3}$P$_{1}$ decay.} \label{repumping}
\end{figure}

If we assume a standard one dimensional model for the MOT we can
determine the influence of the constant $\gamma$. This is a good
approximation as the 881 nm beam is directed along one of the uv MOT
beams, but not retro reflected. Consider atoms on the $-z$ side, see
figure \ref{repumping}. These atoms will mainly populate the $|J=1,
m_j=0,-1 \rangle$ state. Further excitation of the collinear 881 nm
laser will drive the atoms to the left (blue line) or right (red
line) depending on the helicity of the 881 nm laser. This creates an
anisotropy in the decay to the (3s3p)$^{3}$P$_{1}$ state. Atoms
driven along the blue line, i.e., with same polarization helicity as
the MOT beams, will only populate non-trapped states. Experiments
performed in this configurations is thus insensitive to the
re-trapping dynamics as described above. Using the Clebsch Gordan
coefficients shown in figure \ref{repumping} we find the rate
$(1/2\cdot 1 + 1/2\cdot(1/2+1/2))\Gamma_{21}=\Gamma_{21}$, as
expected. For the opposite polarization we find a reduction
$(1/2\cdot 1/2 + 1/2\cdot(2/3+1/6))\Gamma_{21}=2/3\Gamma_{21}$. We
now Consider a more general case where all three magnetic sub states
$|J=1, m_j=+1,0,-1 \rangle$ are populated. The populations are
denoted $n_+$ $n_0$ and with $n_+ + n_0 + n_- =1$. We find the
effective decay constant as $(1-n_+/6)\Gamma_{21}$ for MOT
polarization. The $n_+$ coefficient is expected to be small compared
to the sum of the two other terms. However, even in an extreme case
where all sub states are populated equally $\Gamma_{21}$ is effected
only by about 5\%.

With a 881 nm helicity opposite to that of the MOT beam, marked in
red color on figure \ref{repumping}, will ultimately result in a
destruction of the trapping force as the 881 nm laser intensity
increases. This effect is expected as a result of destructive
reshuffling of the atoms into "wrong" magnetic sub states.

\section{\label{Experiment}Experimental setup and results}

\begin{figure}[h]
\includegraphics*[width=0.95\columnwidth]{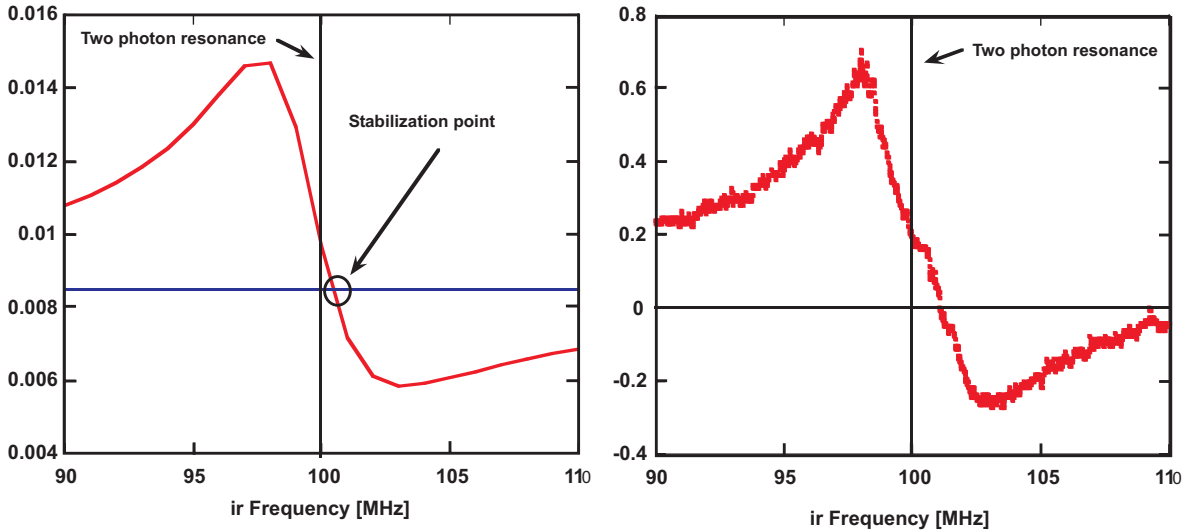}
\caption{ (color online)The 881 nm laser is frequency locked to the
two photon resonance of the ladder system $\nu_{881}=-\nu_{285}=$100
MHz. This ensures a maximal transfer of population to the
$(3s3d)\;^1{\rm D}_2$. Left part of the figure shows the
(3s3p)$^{1}$P$_{1}$ excited state fraction obtained from the optical
Bloch equations. Right part shows the experimental lock
signal.}\label{Lock}
\end{figure}
The main part of the experimental setup has been described in
\cite{Loo}. In our experiment the 285 nm MOT is probed by 881 nm
light produced by a Ti:Sapphire laser. The 881 nm light is power and
frequency controlled by an AOM. A $\lambda$/4 plate controls the
helicity of the light. Presence of the 881 nm light will change the
285 nm light level recorded by a PM (not sensitive to 881 nm light).
This we use to lock the 881 nm laser frequency close to the two
photon resonance $\nu_{881}=-\nu_{285}=$100 MHz during the
experiments, see figure \ref{Lock}. Increasing the rabi frequency of
the 881 nm light causes a minor change in the locking point. In our
power range this corresponds to about 1 MHz and not considered
important at this level of accuracy.

\noindent Typically 10$^7$ atoms are captured in the MOT, with a rms
diameter of 1 mm and temperatures from 3 to 5 mK. The Doppler
cooling limit for the MOT is 2 mK as sub-Doppler cooling is not
supported in this system.

The number of emitted 285 nm photons when the ir laser was on and
off respectively, was measured for each of the two circular
polarizations states. The 881 nm light was sent through the AOM,
which was modulated with a square wave at 1 kHz, with a duty cycle
of 20 \%. The photomultiplier signal was measured and average over
32 periods with the light respectively on and off, thus providing
values for $S^{on}$ and $S^{off}$. At the same time these signals
provided a lock of the Ti:S laser to the two photon resonance, see
figure \ref{Lock}. This measurement was repeated for different ir
intensities (Rabi frequencies), while the MOT was running in steady
state mode.

\begin{figure}
\includegraphics*[width=0.85\columnwidth]{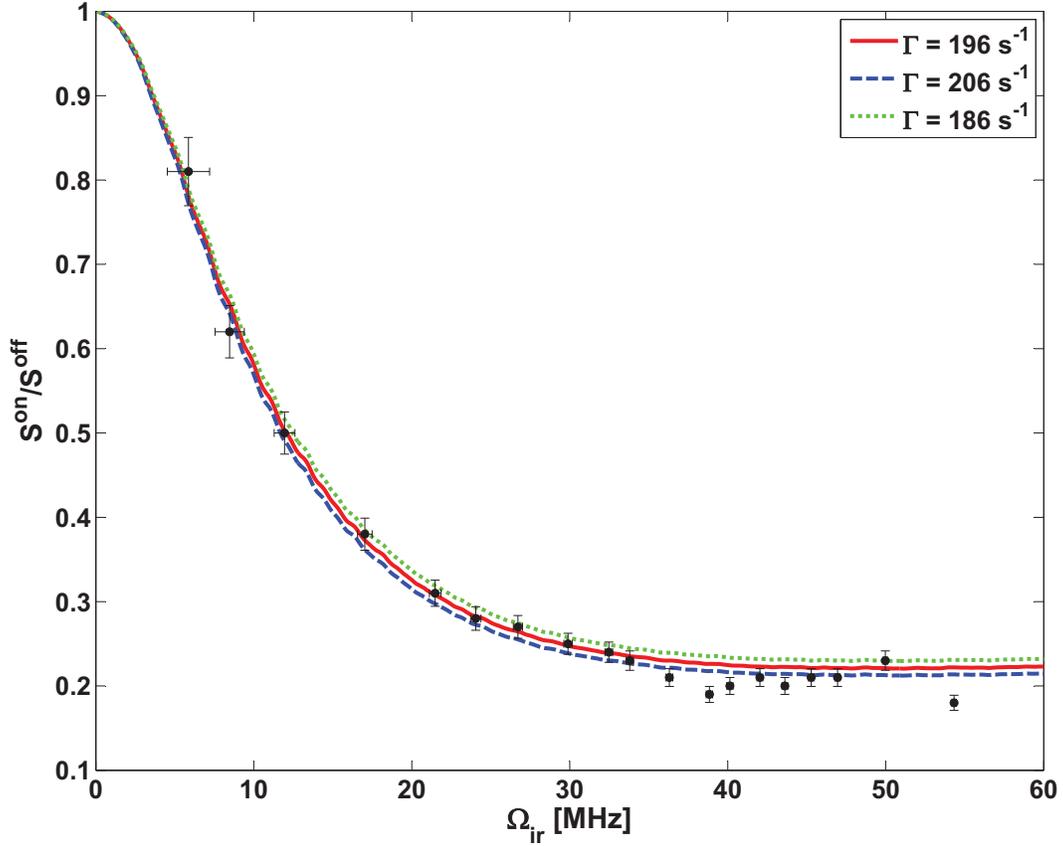}
\includegraphics*[width=0.85\columnwidth]{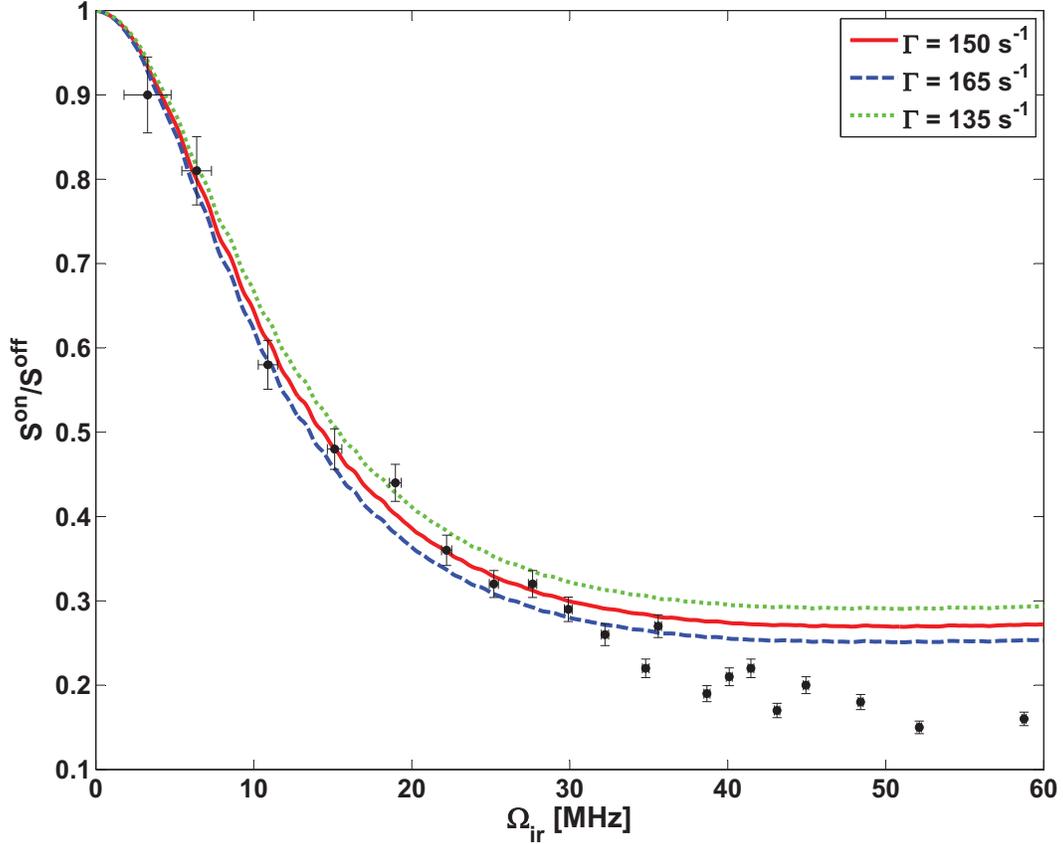}
\caption{(color online) Experimental values for the ratio
$\frac{S^{on}}{S^{off}}$ as a function of the 881 nm rabi frequency.
(top) Here the circular polarization corresponds to the MOT helicity
(blue lines in figure \ref{repumping}). The curves are results from
a simulations based on the optical Bloch equations, including the
magnetic field of the MOT, at different decay constants
$\Gamma=\Gamma_{22}+\Gamma_{21}= 186$ s$^{-1}$, 196 s$^{-1}$, 206
s$^{-1}$. Experiments with this circular helicity supports a value
of $\Gamma=(196\pm10 )$ s$^{-1}$.(bottom) Experiment with 881 nm
polarization opposite to the MOT helicity. At a Rabi frequency of
about 30 MHz the MOT is being destroyed by the destructive optical
pumping effects of the 881 nm laser. Only data points below 30 MHz
should be considered.}\label{pol B}
\end{figure}

Experimental values for the ratio $\frac{S^{on}}{S^{off}}$ as a
function of the 881 nm Rabi frequency are show in figure \ref{pol
B}. As expected the fraction starts out at 1 and gradually decreases
as the Rabi frequency is increased. In figure \ref{pol B} the 881 nm
polarization corresponds to the MOT helicity (blue line in figure
\ref{repumping}) and polarization opposite to the MOT helicity. The
solid curves are results from a simulation based on the optical
Bloch equations including the MOT magnetic field. Experiments with
this helicity clearly supports a value of $\Gamma=(196\pm10 )$
s$^{-1}$ in good agreement with calculations performed in
\cite{porsev}, where $\Gamma=\Gamma_{22}+\Gamma_{21}= (57.3 + 144)$
s$^{-1}$. For the opposite 881 nm polarization (red line in figure
\ref{repumping}) we observe a reduced decay constant of
$\Gamma=(150\pm15 )$ s$^{-1}$. Using values of \cite{porsev} and the
coefficient deduced above we obtain
$\Gamma=\Gamma_{22}+2/3\Gamma_{21}= 153$ s$^{-1}$ in favor of the
one dimensional model described above. From this we find the decay
constants 3s3d$^{1}$D$_{2}$ $\rightarrow$ 3s3p$^{3}$P$_{1}$, and
3s3d$^{1}$D$_{2}$ $\rightarrow$ 3s3p$^{3}$P$_{2}$ to be
$\Gamma_{21}= 138 \pm 10$ s$^{-1}$ and $\Gamma_{22}=58 \pm 4$
s$^{-1}$.

\section{Conclusion}
In conclusion we have measured the total spin forbidden decay rate
(3s3d)$^{1}$D$_{2}$ $\rightarrow$ (3s3p)$^{3}$P$_{2,1}$ in
$^{24}$Mg. We find a rate of $\Gamma=(196\pm10 )$ s$^-1$ supporting
state of the art relativistic many-body calculations. Assuming a
simple one dimensional model for the MOT we find the differential
decay constants (3s3d)$^{1}$D$_{2}$ $\rightarrow$
(3s3p)$^{3}$P$_{1}$, and (3s3d)$^{1}$D$_{2}$ $\rightarrow$
(3s3p)$^{3}$P$_{2}$ as $\Gamma_{21}= 138 \pm 10$ s$^{-1}$ and
$\Gamma_{22}=58 \pm 4$ s$^{-1}$ supporting the above mentioned
calculations.

With a total loss rate of 196 s$^{-1}$ the magnesium MOT lifetime
can be limited to milliseconds time scale or less near the two
photon resonance. To avoid significant atom losses the cooling time
of a two-photon cooling scheme must be kept below 1 ms, as
demonstrated in \cite{Mehlstaubler}. The losses from the $^1$D$_2$
state are according to \cite{Dunn},\cite{Arimondo} limiting the
minimum temperature reachable by two-photon cooling to about 50
$\mu$K. In quantum optics experiments employing this ladder system
decoherence will likewise play a role on relative short timescales.

However the decay rate $\Gamma_{22}$ opens a way of populating the
$^3$P$_2$ state. For our current MOT characteristics a load rate in
the order of 10$^5$ - 10$^6$ atoms/s is expected. Assuming a
magnetic trap lifetime in the order of 20 s, this will result in a
$^3$P$_2$ population of 10$^6$ - 10$^7$ atoms. However, the loading
rate into the $^3$P$_2$ can be increased by increasing the MOT
loading rate using a Zeeman slower. The $^3$P$_2$ state is a fine
starting point for further laser cooling on the $^3$P$_2$ -
$^3$D$_3$ transition as done for Ca \cite{CaMETAMOT}.

\begin{acknowledgments}
We wish to acknowledge the financial support from the Lundbeck Foundation and the Carlsberg Foundation.

\end{acknowledgments}

\end{document}